\documentclass[final,10pt]{svjour2}
\usepackage{graphicx}
\usepackage{amssymb}
\usepackage{amsmath}
\usepackage[dvipsnames]{color}
\usepackage[numbers]{natbib}
\makeatletter
\journalname{Journal of Low Temperature Physics}

\bibpunct{}{}{,}{s}{}{,}

\renewcommand{\vec}[1]{{\mbox{\boldmath$#1$}}} 

\newcommand{\be}{\begin{equation}}
\newcommand{\ee}{\end{equation}}
\newcommand{\ba}{\begin{eqnarray}}
\newcommand{\ea}{\end{eqnarray}}

\newcommand{\chip}{\ensuremath{\chi^2_\mathrm{P}}}
\newcommand{\chin}{\ensuremath{\chi^2_\mathrm{N}}}
\newcommand{\chig}{\ensuremath{\chi^2_\gamma}}
\newcommand{\chigll}{\ensuremath{\chi^2_\mathrm{GMLE}}}
\newcommand{\chimle}{\ensuremath{\chi^2_\mathrm{MLE}}}

\begin{document}

\title{Maximum-likelihood fits to histograms for improved parameter estimation}

\author{
J.~W.~Fowler$^1$
}

\institute{1:National Institute of Standards and Technology, 325
  Broadway MS 686.02, Boulder, CO 80305, USA \\
\email{joe.fowler@nist.gov} \\
\emph{Official contribution of NIST, not subject to copyright in the United States.}
}

\date{22.07.2013}

\maketitle

\begin{abstract}
Straightforward methods for adapting the familiar $\chi^2$ statistic to histograms of discrete events and other Poisson distributed data generally yield biased estimates of the parameters of a model.  The bias can be important even when the total number of events is large.
For the case of estimating a microcalorimeter's energy resolution at 6\,keV from the observed shape of the Mn K$\alpha$ fluorescence spectrum, a poor choice of $\chi^2$ can lead to biases of at least 10\,\% in the estimated resolution when up to thousands of photons are observed.
The best remedy is a Poisson maximum-likelihood fit, through a simple modification of the standard Levenberg-Marquardt algorithm for $\chi^2$ minimization.  Where the modification is not possible, another approach allows iterative approximation of the maximum-likelihood fit.
  
\end{abstract}


\section{Introduction}

One of the most important tasks in experimental physics is estimating the parameters of a model from data.
Often, the data are summarized in the form of a histogram; each bin counts independent, discrete events, and its contents therefore follow the Poisson distribution.\citep{Poisson:1837}  
It is all too easy to make biased parameter estimates by inappropriate use of techniques adapted to counting data from the $\chi^2$ statistic designed for Gaussian distributions.  Fortunately, it is also nearly as easy to use maximum-likelihood estimators, which have substantially less bias---and for certain parameters, none at all.

In the low-temperature detector community, we often assess the performance of a sensor or a multiplexing readout by estimating the energy resolution from measurements of a gamma-ray line (such as the 97 and 103\,keV lines of $^{153}$Gd) or of a fluorescence line (such as the Mn K$\alpha$ line at 5.9\,keV).  This assessment requires fitting a histogram of energies to find the energy resolution, among other parameters.  Clearly, any bias in estimating the resolution can produce misleading results.

Many methods appear in the literature for fitting models to histograms.  Several articles explore their biases in the special case where $N$ independent observations are assumed to measure the same underlying Poisson parameter $\mu$.  Fitting models such as Gaussians to observations is different from fitting a single unknown constant, however, and it is hard to generalize from the published results to the biases on estimates of energy resolution.  The goal of this article is twofold: to establish that maximum-likelihood estimators are the best choice for unbiased estimates of spectral resolution, and to show that such estimators are not inconsistent with fast and convenient computation.


\section{Possible cost functions for fitting models to histograms}

When estimating (``fitting'') parameters of a model from a data set, one generally chooses a cost function $C(\vec{p}; \vec{x})$ that  depends on the model parameters \vec{p} and the observations \vec{x}.  The ``best'' parameters \vec{\hat{p}} are those which minimize $C$ given the data \vec{x}.  For measurements normally distributed with known variances about the (unknown) true values,  the standard $\chi^2$ statistic is the appropriate cost function.  If the variances are uncorrelated with the true values, then minimizing $\chi^2$ also maximizes the likelihood function.

The situation is less clear when the measured data are Poisson distributed, such as the bin contents in a histogram.  The usual $\chi^2$ statistic is a weighted sum of the squared difference between measured and modeled values, with weights equal to the inverse variance of each value.  In Poisson distributed data,  the variance equals the expected value in each measurement, so should the inverse weight be set equal to the measured or the modeled number of events in each bin?  In fact, either choice produces parameter biases.

\begin{table}
%
\vspace{-3mm}
\setlength{\jot}{0pt}
\begin{align}
  \chin &\equiv \sum_{i=1}^{N}\ \frac{(c_i-y_i)^2}{\mathrm{max}(c_i,1)} \label{eq:chin}\\
  \chip &\equiv \sum_{i=1}^{N}\ \frac{(c_i-y_i)^2}{y_i} \label{eq:chip}\\
  \chig &\equiv  \sum_{i=1}^N\, \frac{[c_i + \mathrm{min}(c_i,1) -y_i]^2}{c_i+1} \label{eq:chig} \\
  \chigll &\equiv \chip + \sum_{i=1}^N \log y_i \label{eq:chigll} \\
  \label{eq:chisq_mle}
  \chimle &\equiv -2\log{\cal L}_P = 2\sum_{i=1}^N(y_i-c_i\log\,y_i)
  \end{align}
  
  \caption{Five functions of both data and model that can be minimized with respect to model parameters to yield fits to a histogram.  The last, \chimle, is best for Poisson $c_i$.  It derives from the Poisson-data likelihood function ${\cal L}_P = \prod_{i=1}^N\ \mathrm{e}^{-y_i} y_i^{c_i} / c_i!$, with constant factors (i.e., factors independent of the model \vec{y}) omitted.
  \label{tab:equations}
  }
\vspace{-2mm}
\end{table}

Consider a set of $N$ measurements, the contents of the $N$ bins in a histogram.  Let $c_i$ denote the measured counts in bin $i$.  The parameterized model for the $i$th value is $y(x_i; \vec{p})$, or $y_i$ for short.  If the \vec{c} were normally distributed (which they are not), then the appropriate cost function would be $\chi^2_\mathrm{normal} \equiv \sum_{i=1}^N\, w_i (c_i-y_i)^2 $
with each weight $w_i$ equal to the inverse variance of measurement $c_i$.  Following the convention of Baker and Cousins\cite{Baker:1984p34}, the respective choices of $1/c_i$ or $1/y_i$ as the 
weights for Poisson data are called Neyman's \chin\ and Pearson's \chip\ (Equations~\ref{eq:chin} and \ref{eq:chip}).  In the former case, the weight must be modified to $1/\mathrm{max}(1,c_i)$ to avoid divergences if any bins are empty (any $c_i=0$).  Neyman's \chin\ is the simplest statistic to use in a general-purpose least-squares fitting algorithm, because its weights are independent of the model.  Unfortunately, \chin\ is also the most biased.

Mighell\cite{Mighell:1999p525} advocates another statistic with the practical advantages of \chin, yet constructed specifically to avoid bias at any expected number of counts:
\chig\ (Equation~\ref{eq:chig}).
Hauschild and Jentschel\cite{Hauschild:2001}, however, point out that \chig\ estimators are unbiased only  in the asymptotic limit of  many observations $N$ but are biased for finite $N$. They consider (without advocating) another cost function, the ``Gauss max-likelihood'' (Equation~\ref{eq:chigll}) which improves on the \chip\ in accounting for an additional factor in the Gaussian likelihood function.

The cost function that introduces the least bias is the negative logarithm of the likelihood function for Poisson-distributed data (Table~\ref{tab:equations} and Equation~\ref{eq:chisq_mle}).   This quantity, \chimle, is also known in x-ray astronomy as the Cash $C$-statistic.\cite{Cash:1979p772}  If an additional term is added, a sum over nonzero observations, then the resulting $\chi^2_\lambda\equiv\chimle + 2\sum_{i=0, c_i>0} (c_i\log\,c_i - c_i)  $ is asymptotically distributed\cite{Baker:1984p34} as $\chi^2$ for large $N$.  This term, being independent of the model $y$, does not affect parameter fits.    There is some disagreement in the literature over the best statistic to use in goodness-of-fit tests, with 
some authors\cite{Hauschild:2001} arguing for \chip\ and others\cite{Baker:1984p34,Neyman:1928,Wilks:1935,Fisher:1950} favoring $\chi^2_\lambda$.


\section{Biases on the histogram area and width} \label{sec:biases}

\begin{figure}[tb]
\begin{center}
\parbox[b]{\linewidth}{
  \includegraphics[%
    width=\linewidth,
    keepaspectratio]{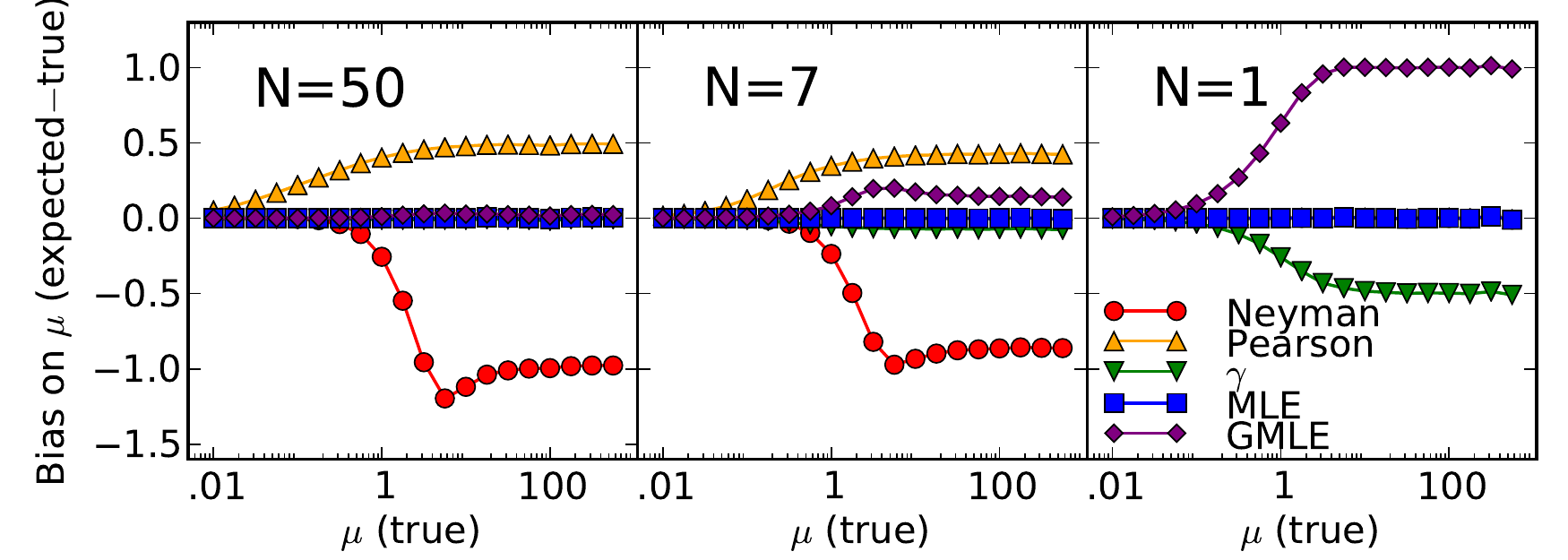}
}
\end{center}
\caption{
The expected additive bias on $\mu$ (expected $-$ true) for five different cost functions.  The model is the trivial case where all data estimate the same constant value of $\mu$.  The three panels have the indicated number $N$ of Poisson variates.  The $N=50$ case is not visibly different from the $N\rightarrow\infty$ limit.
(color figure online) 
}
\label{fig:bias}
\end{figure}

The simplest model to study is that of an $N$-bin histogram with an unknown, constant level: $y_i=\mu$.  In this case, the best-fit estimate $\hat{\mu}$ for all of the cost functions can be expressed in closed form.  The MLE method is the only one that yields the data mean, $\hat{\mu}=(\sum\ c_i)/N$, which is an unbiased estimator of $\mu$.  For the other cost functions, the estimators are:
%
%
%
$ 
\hat{\mu}_N= {N'}/{\sum 1/\mathrm{max}(c_i,1)};
\hat{\mu}_P=\sqrt{\sum c_i^2/N};
\hat{\mu}_\gamma ={N'}/{\sum 1/(c_i+1)};
\mathrm{and}\ 
\hat{\mu}_\mathrm{GMLE}=\sqrt{\hat{\mu}_P + \frac{1}{4}}-\frac{1}{2}$.
Here, $N'$ signifies how many $c_i$ (out of $N$) are non-zero.
From these expressions, it is simple to compute the estimators for simulated Poisson data; the results are shown in Figure~\ref{fig:bias}.  With $N\gtrsim5$ (left and center panels), the Pearson $\hat\mu_P$ has a positive bias as large as $+0.5$ counts when the true value $\mu\gtrsim1$, while the Neyman $\hat\mu_N$ has a negative bias of approximately $-1$ count for $\mu\gtrsim2$.  In the asymptotic limit $N\rightarrow\infty$, the other three estimators are unbiased.  In the opposite limit, where only one value is used to estimate $\mu$, $\hat\mu_P$ and $\hat\mu_N$ are unbiased, while for $\mu\gtrsim 1$, $\hat\mu_\mathrm{GMLE}$ and $\hat\mu_\gamma$ have biases of $+1$ and $-0.5$ counts.  

For more complicated model functions, these results generalize readily if the model contains a parameter (or combination of them) that allows the entire model to scale freely by a  factor.
The new statement is that \chin\ and \chip\ produce biased estimators of a histogram's area when more than a few bins are used, and \chig\ and \chigll\ produce biased area estimators when not many bins are used.  Only the MLE value is unbiased for any area and all $N$.
In some circumstances, biases on the fitted area might be unimportant.  Still, using the one estimator mathematically guaranteed not to bias the area seems prudent in any conditions.

\begin{figure}[tb]
\begin{center}
\parbox[b]{\linewidth}{
  \includegraphics[%
    width=\linewidth,
    keepaspectratio]{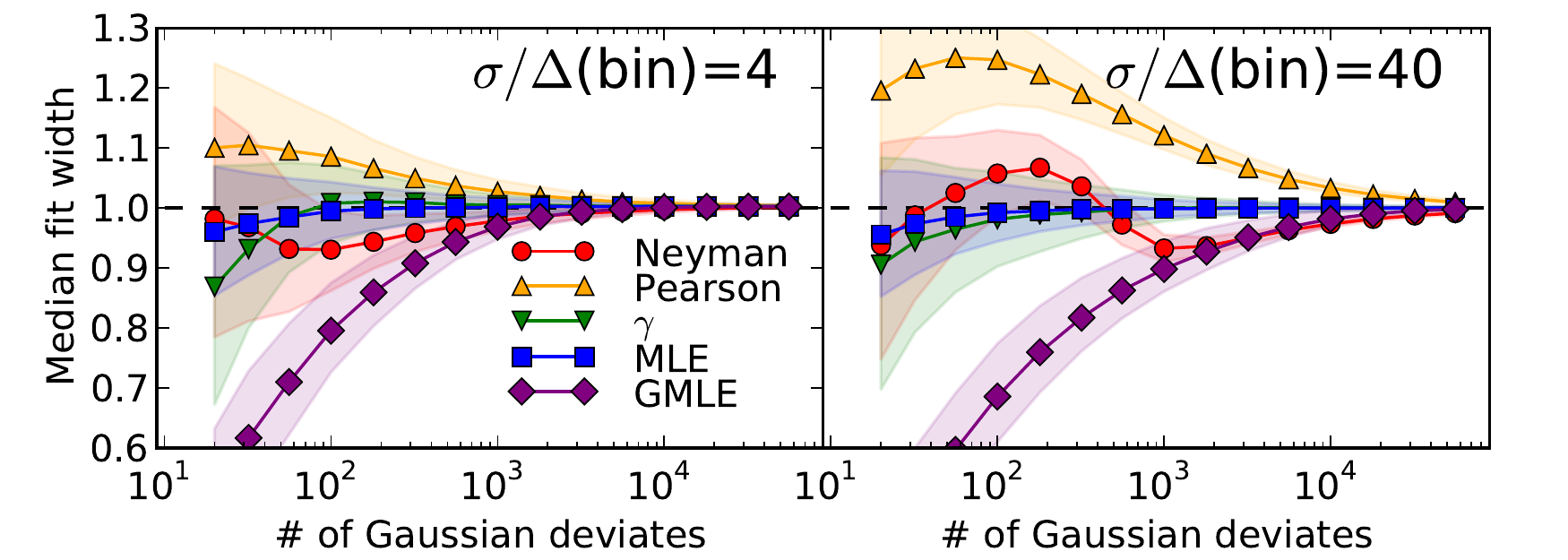}
}
\end{center}
\caption{Median width in fit from unit-variance Gaussian deviates histogrammed with wide and narrow bin widths $w=\sigma/4$ and $w=\sigma/40$.  Shaded bands contain the middle two quartiles of the 8000 fits per point.
The simulated data have no constant background, and none is fit.  In all cases, the bias is largest if the total number of events $N_e$ is small and if narrow bins (\emph{right panel}) produce a small number of events per bin.  The Poisson Maximum Likelihood (``MLE'') fit is the least biased of the methods, though it shows a small bias at $\lesssim100$ events.  Mighell's \chig\ is also acceptable for $\gtrsim 10$ events per bin at the peak.
(color figure online) 
}
\label{fig:gauss}
\end{figure}

The width of a spectral peak is another parameter often estimated from Poisson-distributed histogram data.  Figure~\ref{fig:gauss} shows the median $\hat\sigma$ fit to simulated Gaussian deviates  with 4 or 40 bins per $\sigma$.  The most important result is that the \chig\ and \chimle\ estimators are the only ones without significant bias on $\hat\sigma$ for fits to  $\gtrsim100$ Gaussian deviates.  Below 100 events (in either binning), a small underestimate in width is apparent.  In such conditions of small peaks, Monte Carlo modeling of the fit is probably necessary.

\begin{figure}[tb]
\begin{center}
\parbox[b]{\linewidth}{
  \includegraphics[%
    width=\linewidth,
    keepaspectratio]{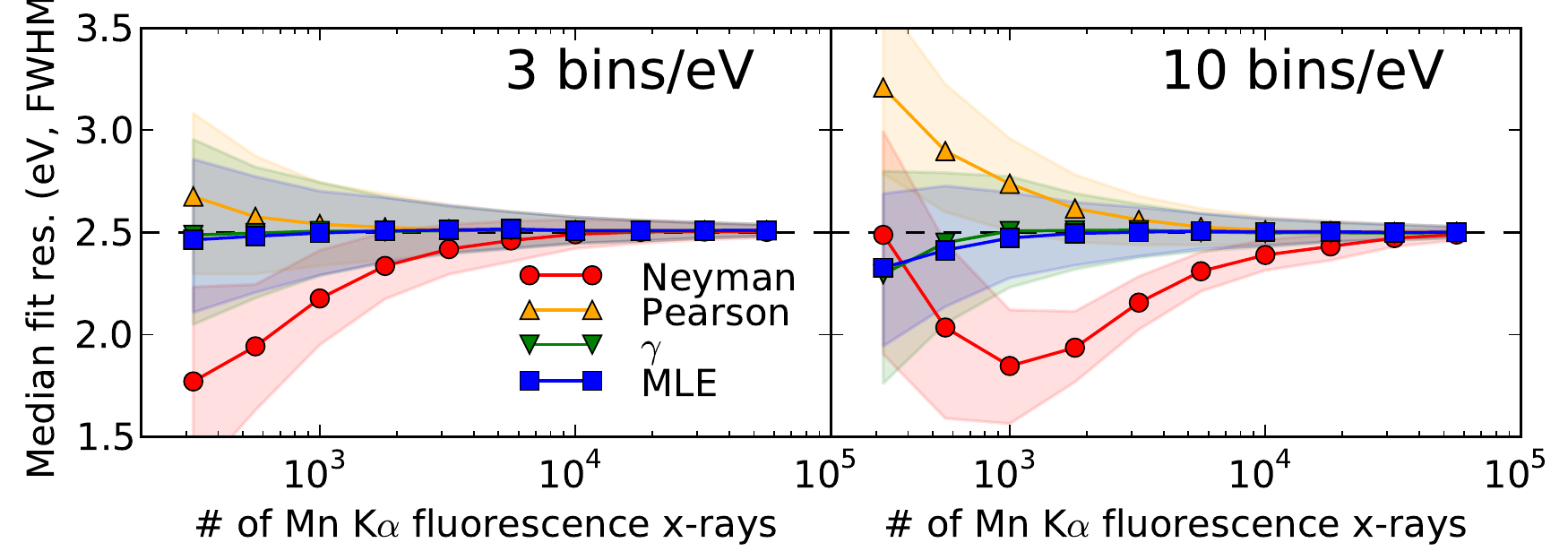}
}
\end{center}
\caption{
The median fit result for energy resolution using the Mn~K$\alpha$ complex, with true energy resolution 2.5\,eV (FWHM).   Shaded bands contain the middle two quartiles of the 3000 fits per point. The data are binned with 3 and 10 bins per eV, putting approximately 4\% or 1\% of all values in the histogram's peak bin.  A constant background per bin varies freely in the fit, but the true value is 0.  The usual Neyman \chin\ severely underestimates the energy resolution, while Pearson's overestimates it.  Achieving low bias is possible with the other \chig\ and MLE cost functions, given $\gtrsim 10$ counts per bin at the  peak.
(color figure online) 
}
\label{fig:mnkalpha}
\end{figure}

Figure~\ref{fig:mnkalpha} shows the median Gaussian energy resolution (FWHM) estimated by fitting the Mn~K$\alpha$ x-ray fluorescence spectrum\cite{Holzer:1997} to simulated energies in the range 5860 to 5930\,eV\@.  Each point is the result of fitting 3000 simulations.  The true resolution is 2.5\,eV\@.  As with fits to the width of  Gaussian distributions, the \chin\ and \chip\ estimators exhibit substantial bias and cannot be recommended. (The same is true of \chigll\ though its minimizer often failed to converge, and it is not shown here.)
In this estimation problem, both \chimle\ and \chig\ lead to minimal bias, though care is again required with spectra containing fewer than $\sim1000$ photons.

In these tests of fitting parameters related to the area under a histogram peak and to its width, the biases of the earliest and simplest methods (\chin\ and \chip) are manifest.  The \chig\ approach seems to have low bias on the width parameters, but  the area bias in Figure~\ref{fig:bias} is a cause for concern.  Only the maximum-likelihood estimators perform well on all bias tests.

\section{Efficient maximum-likelihood parameter estimation}
If the maximum-likelihood estimators are preferred, how can they be computed efficiently?  Some software  computes them without modification; for example, the CERN library MINUIT.  More often, software libraries or packages offer only the extremes of fully general nonlinear minimization (which can be slow or fragile, or both) and ``least-squares minimization.'' Routines of the latter sort can find the minimum of a $\chi^2$ with constant (i.e., model-independent) weights, as in Equations~\ref{eq:chin} or \ref{eq:chig}. Examples include \verb=gsl_multifit_fdfsolver= \verb=_lmsder= in gsl, \verb=Fitmrq= in \emph{Numerical Recipes},\cite{Numrec} \verb=scipy.optimize.leastsq= in Python, and \verb=leasqr= in Octave. They assume that the nonlinear function $\chi^2(y(\vec{p}))$ is strictly quadratic in the set of predicted values $\{y_i\}$, while $y_i(\vec{p})$ may be a general nonlinear function.  An effective algorithm for least-squares fitting is the Levenberg-Marquardt (LM) method,\cite{Levenberg:1944, Marquardt:1963p518} which can outperform general nonlinear minimizers by exploiting its knowledge that the target function is a sum of squares.  A least-squares fitter is not appropriate for minimizing a function like \chimle, which is not quadratic in $y_i$.  Fortunately, it is possible to find maximum-likelihood solutions either by using a slightly modified LM-based fitter or by iterating the results generated by a standard least-squares fitter.


\emph{Modified Levenberg-Marquardt method}
The LM algorithm is commonly identified as a least-squares fitter because it is so widely used for this one purpose. In fact, LM is readily adapted to minimize any composite function $f(y(\vec{p}))$ even when $f$ is not quadratic in $y$, provided that we can compute the first and second  derivatives of $f$.  Using the labels  and scaling of \emph{Numerical Recipes}, for the Poisson maximum-likelihood $f=\chimle$, these are:
\begin{align} 
\label{eq:beta}
\beta_j &= -\frac{1}{2} \frac{\partial f}{\partial p_j} =   \sum_{i=1}^N\,\left(\frac{c_i}{y_i}-1\right)\frac{\partial y_i}{\partial p_j}, \\
 \label{eq:alpha}
\alpha_{jk} &= \frac{1}{2}\frac{\partial^2 f}{\partial p_j\,\partial p_k} \approx   \sum_{i=1}^N\, \left(\frac{c_i}{y_i^2} \right)\frac{\partial y_i}{\partial p_j} \frac{\partial y_i}{\partial p_k}.
\end{align}
The equality is only approximate because terms containing second derivatives of $y_i$ are dropped; they generally have a destabilizing or minimal effect on the LM performance.  Adapting the standard LM to achieve a Poisson maximum-likelihood fit instead requires only three changes: using the parenthesized factors  in Equations~\ref{eq:beta} and \ref{eq:alpha} in place of $[w_i(c_i-y_i)]$ and $[w_i]$ respectively in the standard $\chi^2$ minimization, and computing \chimle\ (Equation~\ref{eq:chisq_mle}) in place of $\chi^2$. When starting from source code that implements the basic LM least-squares algorithm, these replacements should be straightforward.\cite{Laurence:2010p144} This is the approach used for the fits summarized in Section~\ref{sec:biases}.

Incidentally, by setting $\beta_0=0$ in Equation~\ref{eq:beta}, it is easy to show that a model of the form $y=p_0 f_i(p_1,\dots p_m)$ guarantees $\sum_i\,c_i = \sum_i\,y_i$.  That is, a model having an overall scale factor as one parameter is guaranteed to have equal observed and predicted total counts in a maximum-likelihood fit.\cite{Baker:1984p34}


\emph{Iterative least-squares method:}
It is also possible to approach the Poisson maximum-likelihood fit using only the framework of a standard LM least-squares fitter by employing an additional ``outer iteration'' over least-squares fits.  
Compare the desired MLE solution (set Equation~\ref{eq:beta} to zero) with
\be \label{eq:grad_chisq_norm}
0 = \frac{\partial \chi^2_\mathrm{normal}}{\partial p_j} = \sum_{i=1}^N\,w_i(c_i-y_i)\frac{\partial y_i}{\partial p_j}.
\ee
If the predicted counts $\hat{y}_i$ arise from the maximum-likelihood model, then by definition they solve Equation~\ref{eq:beta}.  The same model solves Equation~\ref{eq:grad_chisq_norm} if the weights are taken to be $w_i\equiv 1/\hat{y}_i$.  This suggests an iterative approach to solving the desired Equation~\ref{eq:beta} employing a least-squares solver capable only of solving Equation~\ref{eq:grad_chisq_norm}.  Start by minimizing \chin, i.e., solve Equation~\ref{eq:grad_chisq_norm} with $w_i=\mathrm{max}(1,c_i)$, to obtain an initial guess at the model parameters $\vec{p}_0$.  Use constant weights $w_i=1/y_i(\vec{p}_0)$ to solve the least-squares equation again.  This produces an improved set of parameters $\vec{p}_1$ and weights, from which another iteration can begin.  This method appears to be discussed first by Wheaton et al.\cite{Wheaton:1995p519} In at least some conditions, two rounds of least-squares fits  suffice to remove the bias introduced in the initial fit that minimized \chin.

\section{Conclusion}

Two practical approaches were given to maximize the Poisson likelihood and minimize \chimle.  It is important to do this rather than to minimize Neyman's \chin\ when estimating energy resolution from a spectrum, or the estimate will give an overly optimistic view of detector performance, with a bias comparable to the uncertainty on the resolution.  Minimizing Pearson's \chip\ has the opposite effect. Estimates of shape parameters from a data histogram are best found through maximum-likelihood fits, which have lower bias than other approaches and are not difficult to perform.

\begin{acknowledgements}
The author was supported by an American Recovery and Reinvestment Act senior fellowship and by the NIST Innovations in Measurement Science program.  The author thanks J\@. Ullom for encouragement and many helpful discussions and C\@. Pryke for debates on the topic long ago.
\end{acknowledgements}

\newcommand{\apl}[1]{\emph{Appl.\ Phys.\ Lett.\ }{\bf #1}}

\end{document}